\documentstyle[mprocl]{article}

\input{psfig}
\def\be{\begin{equation}}
\def\ee{\end{equation}}
\def\bea{\begin{eqnarray}}
\def\eea{\end{eqnarray}}

\begin{document}

\title{Anisotropy in the Compressible Quantum Hall State}

\author{Nobuki Maeda}

\address{Department of Physics, Hokkaido University, 
Sapporo 060-0810, Japan\\E-mail: maeda@particle.sci.hokudai.ac.jp}   


\maketitle\abstracts{
Using a mean field theory on the von Neumann lattice, we study 
compressible anisotropic states around $\nu=l+1/2$ in the 
quantum Hall system. The Hartree-Fock energy of 
the UCDW are calculated self-consistently. 
In these states the unidirectional charge density wave (UCDW) 
seems to be the most plausible state. 
We show that the UCDW is regarded as a collection of the one-dimensional 
lattice fermion systems which extend to the uniform direction. 
The kinetic energy of this one-dimensional system is induced from 
the Coulomb interaction term and the self-consistent Fermi surface 
is obtained. }

\section{Introduction}
As summarized in the table below, the physics at the half-filled 
$l$-th Landau level is drastically changed in the different Landau 
level space. 
In particular, the nature of the anisotropic states discovered 
in the third or higher Landau levels are unknown until now. 
At the half-filled lowest and second Landau levels, transition to the 
anisotropic state was also observed in the presence of the periodic 
potential or in-plane magnetic field, respectively. 

\begin{figure}[b]
\psfig{figure=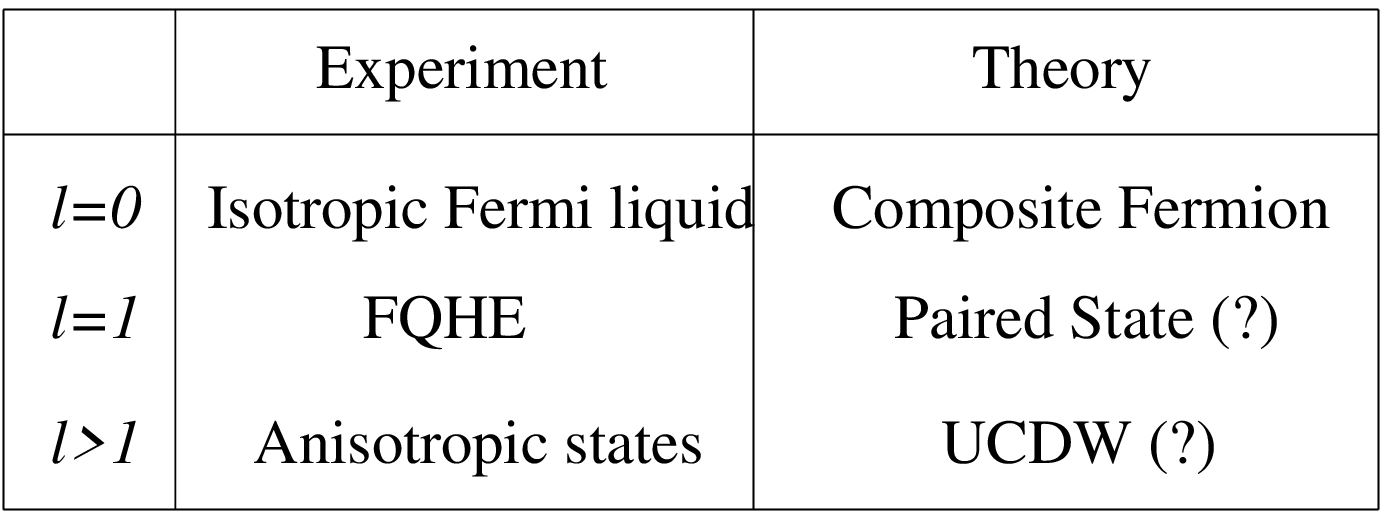,height=1.8in}
\end{figure}

The charge density wave state (CDW)~\cite{c,d,e,f,g} is a candidate 
of the anisotropic state. 
We study the compressible charge density wave (CCDW) states, 
which has no energy gap, 
by the Hartree-Fock approximation using the 
von Neumann lattice formalism.~\cite{b,a} 

There are two types of The CCDW states, the one is the unidirectional 
charge density wave (UCDW) state and the other is the compressible 
Wigner crystal (CWC) state. 
The UCDW state has a charge density which is uniform in one direction 
and oscillates in the other direction. 
The charge density of the CWC state has the same periodicity of the 
von Neumann lattice. 
The classification of the CDW is given in the next table. 
\begin{figure}[b]
\psfig{figure=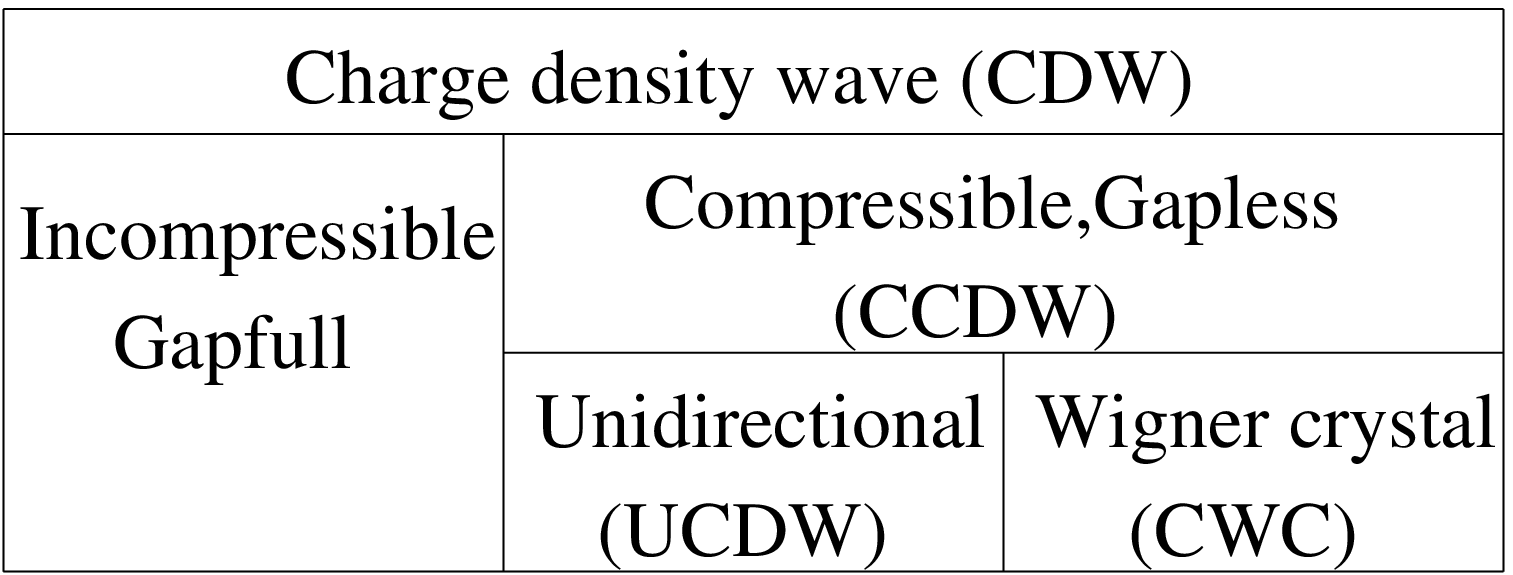,height=1.8in}
\end{figure}

\section{Mean Field Theory on the von Neumann Lattice}
Let us consider the two-dimensional 
electron system in a perpendicular magnetic field $B$ which is described 
by a Hamiltonian $H=H_0+H_{\rm int}$, 
\begin{eqnarray}
H_0&=&\int d^2 r\psi^\dagger({\bf r}){({\bf p}+e{\bf A})^2\over2m}
\psi({\bf r}),\\
H_{\rm int}&=&{1\over2}\int d^2 rd^2 r':\rho({\bf r})V({\bf r}-{\bf r}')
\rho({\bf r}'):.
\end{eqnarray}
where $\rho({\bf r})=\psi^\dagger({\bf r})\psi({\bf r})$, 
$\nabla\times{\bf A}=B$, and $V({\bf r})=q^2/{\rm r}$. 

In the von Neumann lattice formalism,~\cite{b,a} 
the electron field is expanded as 
$\psi({\bf r})=\sum_{l,{\bf X}}b_l({\bf X})W_{l,{\bf X}}({\bf r})$, 
where $b$ is an anti-commuting annihilation operator and $\bf X$ is 
an integer valued two-dimensional coordinate. 
The Wannier basis $W_{l,{\bf X}}({\bf r})$'s are orthonormal 
complete basis in the $l$-th Landau level as 
$(W_{l,{\bf X}}({\bf r}),W_{l',{\bf X}'}({\bf r}))=
\delta_{ll'}\delta_{{\bf X},{\bf X}'}$. 
$W_{l,{\bf X}}({\bf r})$ are
localized at two-dimensional lattice sites $m{\bf e}_1+n{\bf e}_2$ for 
${\bf X}=(m,n)$, where ${\bf e}_1=(ra,0)$, ${\bf e}_2=(a/r\tan\theta,
a/r)$, and $a=\sqrt{2\pi/e B}$. 
The area of the unit cell is ${\bf e}_1\times{\bf e}_2=a^2$. 
We set $a=1$ in the following calculation. 

The Bloch wave basis $u_{l,{\bf p}}({\bf r})=
\sum_{\bf X}W_{l,{\bf X}}({\bf r})e^{i{\bf p}\cdot{\bf X}}$ is 
another useful basis. 
We can obtain the charge density profile~\cite{a} of the CDW state 
using this basis. 
The lattice momentum $\bf p$ is defined in the Brillouin zone (BZ), 
$\vert p_i\vert\leq \pi$. 
Using this basis, we obtain another expansion of the electron field as 
$\psi({\bf r})=\sum_l\int_{\rm BZ}{d^2p\over(2\pi)^2}
a_l({\bf p})u_{l,{\bf p}}({\bf r})$. 
The Fourier transformed density operator 
$\tilde\rho({\bf k})=\int d^2r\rho({\bf r})e^{i{\bf k}\cdot{\bf r}}$ 
is written as
\begin{equation}
\tilde\rho(\tilde{\bf k})=\sum_{l,l'}\int_{\rm BZ}{d^2p\over(2\pi)^2}
a^\dagger_l({\bf p}+{\bf k})M_{ll'}({\bf k})e^{if({\bf p}+{\bf k},
{\bf p})}a_{l'}({\bf p}),
\label{rho}
\end{equation}
where $\tilde{\bf k}=(k_x/r,rk_y-k_x/r\tan\theta)$. 
The phase function $f$ is given by 
$f({\bf p}',{\bf p})
=\int_{{\bf p}}^{{\bf p}'}(\tilde{\bf A}({\bf p})
-\nabla\lambda({\bf p}))\cdot d{\bf p}$, 
where $\tilde{\bf A}({\bf p})=(p_y/2\pi,0)$, which represents a uniform 
magnetic field in the momentum space $\nabla_p\times{\tilde{\bf A}}
({\bf p})=-1/2\pi$. 
The following boundary condition is required, 
$e^{i\lambda({\bf p}+2\pi{\bf N})-i\lambda({\bf p})}
=(-1)^{N_x+N_y}e^{-iN_yp_x},\ 
N_x,\ N_y:\ {\rm integers}$. 
The matrix $M_{l,l'}$ is given by
$\left({l\over l'}\right)^{1\over2}
\left({k_x+ik_y\over\sqrt{4\pi}}\right)^{l'-l}
L_l^{(l'-l)}({k^2\over4\pi})e^{-{k^2\over8\pi}}$, 
for $l\leq l'$ and $M_{l'l}({\bf k})=M^*_{ll'}(-{\bf k})$. 

The free Hamiltonian $H_0$ and interaction 
Hamiltonian $H_{\rm int}$ become
\begin{eqnarray}
H_0&=&\sum_{l,{\rm X}}\omega_c(l+{1\over2})b_l^\dagger({\bf X})
b_l({\bf X}),\\
H_{\rm int}&=&{1\over2}\sum_{{\bf X}_i,l_i,l'_i}:
b_{l_1}^\dagger({\bf X}_1)b_{l'_1}({\bf X}'_1)
V_{l_1 l'_1 l_2 l'_2}({\bf X},{\bf Y},{\bf Z})
b_{l_2}^\dagger({\bf X}_2)b_{l'_2}({\bf X}'_2):
\nonumber
\end{eqnarray}
where ${\bf X}={\bf X}_1-{\bf X}'_1$, 
${\bf Y}={\bf X}_2-{\bf X}'_2$, and 
${\bf Z}={\bf X}_1-{\bf X}'_2$. 
Thus the system is translationally invariant on the lattice. 
${\tilde V}({\bf k})=2\pi q^2/k$ for ${\bf k}\neq0$ and 
${\tilde V}(0)=0$ due to the charge neutrality condition. 

The Hamiltonian $H$ is invariant under the transformation 
$b_l({\bf X})\rightarrow e^{i{\bf K}\cdot{\bf X}} b_l({\bf X})$, 
$\lambda({\bf p})\rightarrow
\lambda({\bf p}+{\bf K})+K_x p_y$. 
This transformation is the magnetic translation in the 
momentum space ${\bf p}\rightarrow{\bf p}+{\bf K}$. 
This invariance is referred to as the K-invariance in the composite 
fermion model. 

\section{Hartree-Fock energy for the CCDW states}
We consider only the intra-Landau level's energy of the $l$-th Landau 
level. 
The filling factor $\nu$ is written as 
$\nu=l+\bar\nu$.
Mean field $U_l({\bf X}',{\bf X})=
\langle b^\dagger_l({\bf X})b_l({\bf X}')\rangle$ for the CCDW 
which has the translational 
invariance on the von Neumann lattice, that is, 
$U_l({\bf X}-{\bf X}')=U_l({\bf X},{\bf X}'),\ 
U_l(0)=\bar\nu$.
The Hartree-Fock Hamiltonian in the $l$-th Landau level, then, becomes 
\begin{equation}
H ^{(l)}_{\rm HF}=\sum_{{\bf X},{\bf X}'} U_l({\bf X}-{\bf X}')\{
{\tilde v}_l (2\pi ({\hat{\bf X}}-{\hat{\bf X}}'))-v_l({\hat{\bf X}}-
{\hat{\bf X}}')\}
\{b^\dagger_l({\bf X}) b_l ({\bf X}')-
{1\over2} U_l({\bf X}'-{\bf X})\}.
\label{hart}
\end{equation}
where 
${\tilde v}_l ({\bf k})=\{L_l({k^2 \over 4\pi})\}^2e^{-{k^2 \over 4\pi}} 
{\tilde V}({\bf k})$, 
$v_l ({\bf X})=\int{d^2k\over (2\pi)^2}{\tilde v}_l ({\bf k}) e^{i{\bf k}
\cdot{\bf X}}$, 
and $\hat{\bf X}=(rm+n/r\tan\theta,n/r)$ for ${\bf X}=(m,n)$. 
Thus the continuum system with a magnetic field is transformed 
to the lattice system without a magnetic field!

The self-consistency equations for the kinetic energy $\varepsilon_l$ is 
given by 
$\varepsilon_l({\bf p},\bar\nu)=\int_{\rm BZ}{d^2 p'\over (2\pi)^2} 
{\tilde v}^{\rm HF}_l({\bf p}'-{\bf p})\theta(\mu_l-
\varepsilon_l({\bf p}',\bar\nu))$, 
where $\mu_l$ is the chemical potential and ${\tilde v}^{\rm HF}_l$ 
is defined by 
${\tilde v}^{\rm HF}_l({\bf p})=\sum_{\bf X}\{
{\tilde v}_l (2\pi ({\hat{\bf X}}))-v_l({\hat{\bf X}})\}e^{-i{\bf p}
\cdot{\bf X}}$. 
The energy per particle in the $l$-th Landau level is given by 
$E^ {(l)}={1\over2\bar\nu}\sum_{\rm X}\vert U_l({\bf X})
\vert^2\{ {\tilde v}_l(2\pi\hat{\bf X})-v_l(\hat{\bf X})\}$. 
$E^{(l)}$ is a function of $\bar\nu$, $r$ and $\theta$. 
The parameters $r$ and $\theta$ are determined so as to 
minimize the energy $E^{(l)}$ at a fixed $\bar\nu$. 
Existence of a Fermi surface breaks the K-invariance inevitably. 
There are two types of self-consistent Fermi seas.

\noindent
(a) {\it\bf Belt-shaped Fermi sea} (UCDW state) : 
\begin{equation}
U_l({\bf X})=\delta_{m,0}{\sin(p_{\rm F}n)\over \pi n},\ 
p_{\rm F}=\pi\bar\nu. 
\end{equation}

\noindent
(b) {\it\bf Diamond-shaped Fermi sea} (CWC state for $\bar\nu=1/2$) : 
\begin{equation}
U_l({\bf X})={2\over(\pi)^2}{\sin{\pi\over2}(m+n)\sin{\pi\over2}(m-n)
\over m^2-n^2}.
\end{equation}

We calculated the Hartree-Fock energy for (a) and (b) at $l<4$. 
As a result, we found that 
the UCDW state is the lowest energy state 
in all cases.~\cite{a} 
Therefore the UCDW state is the most plausible state in the 
CCDW states. 

Using the mean field of (a), the kinetic term in 
$H^{(l)}_{\rm HF}$ is written as 
\begin{equation}
K^{(l)}_{\rm HF}=\sum_m\int{dp_y\over 2\pi} 
a^\dagger_{l,m}(p_y)\varepsilon_l(p_y,\bar\nu)a_{l,m}(p_y),
\end{equation}
where $a_{l,m}(p_y)=\sum_n b_l({\bf X})e^{-ip_y n}$ for ${\bf X}=(m,n)$. 
Therefore the UCDW state is regarded as a collection of the 
one-dimensional lattice Fermi-gas systems which extend to the 
y-direction. 

Using the Buttiker-Landauer formula, the conductance of the UCDW
$\sigma_{xx}=0,$ $\sigma_{yy}=n_x {e^2\over 2\pi}$, 
where $n_x$ is a number of the one-dimensional channels. 
If we take $\sigma_{xy}=\nu e^2/2\pi$, the resistance becomes 
$\rho_{xx}={n_x\over\nu^2}{2\pi\over e^2}$, 
$\rho_{yy}=0$. 
Thus the formation of the UCDW leads the anisotropy in the 
magnetoresistance.

\section{Summary and discussion}

We have studied the CCDW state, which is gapless state 
and has an anisotropic Fermi surface. 
We obtained two types of the CCDW state, the UCDW state and CWC state.  
By calculating the Hartree-Fock energy, the UCDW is found to have a lower 
energy at the half-filled Landau levels. 
The UCDW state is regarded as a system which consists of 
many one-dimensional lattice 
Fermi-gas systems which extend to the uniform direction. 
Formation of this structure could be the origin of the anisotropy 
observed in experiments. Theoretical works to 
include fluctuations around the mean field solution are necessary. 
Since there is no energy gap in the CCDW state, the fluctuation 
effect might be large compared with the gapfull CDW state.

\eject


\vspace*{-9pt}

\section*{References}
\begin{thebibliography}{99}
\bibitem{c}A. A. Koulakov, M. M. Fogler, and B. I. Shklovskii, Phys. Rev. 
Lett. {\bf 76}, 499 (1996); Phys. Rev. B {\bf 54}, 1853 (1996). 
\bibitem{d}R. Moessner and J. T. Chalker, Phys. Rev. B {\bf 54}, 5006 (1996). 
\bibitem{e}E. Fradkin and S. A. Kivelson, Phys. Rev. B {\bf 59} 8065 (1999). 
\bibitem{f}E. H. Rezayi, F. D. M. Haldane, and K. Yang, Phys. Rev. Lett. 
{\bf 83}, 1219 (1999).
\bibitem{g}T. Jungwirth, A. H. MacDonald, L. Smr{\v c}ka, and S. M. Girvin, 
cond-mat/9905353.
\bibitem{b}K. Ishikawa, N. Maeda, and T. Ochiai, Phys. Rev. Lett. 
{\bf 82}, 4292 (1999).
\bibitem{a}N. Maeda, cond-mat/9908373.
\end{thebibliography}
\end{document}